\documentclass[prl, twocolumn,floatfix,superscriptaddress,citeautoscript]{revtex4-2}
\usepackage{graphicx}
\usepackage{dcolumn}   
\usepackage{bm}        
\usepackage{amssymb}  
\usepackage{verbatim}
\usepackage{multirow}
\usepackage{xcolor}
\usepackage{amsmath}
\usepackage[normalem]{ulem}

\usepackage{amsfonts}
\usepackage{hyperref}
\usepackage[normalem]{ulem}   
\hypersetup{colorlinks=true, pdfstartview=FitV, linkcolor=blue, citecolor=black, plainpages=false, pdfpagelabels=true, urlcolor=blue}
\usepackage[all]{hypcap}
\usepackage{physics}
\usepackage{float}

\newcommand{\Gup}{\Gamma_{\uparrow}}
\newcommand{\Gdown}{\Gamma_{\downarrow}}
\newcommand{\gqtls}{g_{\rm{q-TLS}}}
\newcommand{\tenv}{T_{1,\rm{Env}}}
\newcommand{\ttls}{T_{1,\rm{TLS}}}
\newcommand{\tq}{T_{1,\rm{q}}}

\begin{document}

\title{Quantum environment afterglow from broadband excitation spectroscopy in superconducting qubits}
\author{J. R. Guimarães}
\email[]{j.guimaraes@fz-juelich.de}
\thanks{These authors contributed equally}
\affiliation{Institute for Functional Quantum System (PGI-13), Forschungszentrum Jülich, 52425 Jülich, Germany}
\affiliation{Department of Physics, RWTH Aachen University, 52074 Aachen, Germany}

\author{Y.~Liu}
\email[]{ye.liu@fz-juelich.de}
\thanks{These authors contributed equally}
\affiliation{Institute for Functional Quantum System (PGI-13), Forschungszentrum Jülich, 52425 Jülich, Germany}

\author{Y.~Gao}
\affiliation{Institute for Functional Quantum System (PGI-13), Forschungszentrum Jülich, 52425 Jülich, Germany}
\affiliation{Department of Physics, RWTH Aachen University, 52074 Aachen, Germany}

\author{Y.~Haddad}
\affiliation{Institute for Functional Quantum System (PGI-13), Forschungszentrum Jülich, 52425 Jülich, Germany}
\affiliation{Department of Physics, RWTH Aachen University, 52074 Aachen, Germany}

\author{A.~Galicia}
\affiliation{Institute for Functional Quantum System (PGI-13), Forschungszentrum Jülich, 52425 Jülich, Germany}
\affiliation{Department of Physics, RWTH Aachen University, 52074 Aachen, Germany}

\author{D.~A.~Volkov}
\affiliation{Institute for Functional Quantum System (PGI-13), Forschungszentrum Jülich, 52425 Jülich, Germany}
\affiliation{Department of Physics, RWTH Aachen University, 52074 Aachen, Germany}

\author{J.~T.~Schmieder}
\affiliation{Institute for Functional Quantum System (PGI-13), Forschungszentrum Jülich, 52425 Jülich, Germany}
\affiliation{Department of Physics, RWTH Aachen University, 52074 Aachen, Germany}

\author{H.~Bhardwaj}
\affiliation{Institute for Functional Quantum System (PGI-13), Forschungszentrum Jülich, 52425 Jülich, Germany}
\affiliation{Department of Physics, RWTH Aachen University, 52074 Aachen, Germany}

\author{M.~Neis}
\affiliation{Institute for Functional Quantum System (PGI-13), Forschungszentrum Jülich, 52425 Jülich, Germany}
\affiliation{Department of Physics, RWTH Aachen University, 52074 Aachen, Germany}

\author{J.~Cereijo}
\affiliation{Institute for Functional Quantum System (PGI-13), Forschungszentrum Jülich, 52425 Jülich, Germany}
\affiliation{Department of Physics, RWTH Aachen University, 52074 Aachen, Germany}

\author{M.~Jerger}
\affiliation{Institute for Functional Quantum System (PGI-13), Forschungszentrum Jülich, 52425 Jülich, Germany}

\author{P.~A.~Bushev}
\affiliation{Institute for Functional Quantum System (PGI-13), Forschungszentrum Jülich, 52425 Jülich, Germany}

\author{R.~Barends}
\email[]{r.barends@fz-juelich.de}
\affiliation{Institute for Functional Quantum System (PGI-13), Forschungszentrum Jülich, 52425 Jülich, Germany}
\affiliation{Department of Physics, RWTH Aachen University, 52074 Aachen, Germany}
\date{\today}
\begin{abstract}

Understanding the qubit environment is central to fault-tolerant superconducting quantum computation. Characterization typically relies on relaxation, leaving excitation largely underexplored. Here, we access this information with time-resolved broadband excitation spectroscopy. The resulting qubit excitation spectrum reveals a highly structured landscape, interspersed with cold regions. Combined with postselection, this technique enables full reconstruction of the noise power spectral density (PSD) and separates quantum from classical noise. With feed-forward, it exposes long-lived two-level-systems (TLSs), whose relaxation times span tens of microseconds to milliseconds - uncovering an intrinsic link between the qubit-TLS coupling and TLS relaxation. The data suggest that the long-lived TLSs are intrinsic to the qubit environment, and can cause excitation that lasts for many qubit operation cycles. Consequently, the environment retains a memory of prior dynamics, and properties like gate fidelity become non-Markovian and protocol-dependent. The presented approach enables identifying and bypassing the hidden roadblocks formed by long-lived TLSs in fault-tolerant quantum computation.
\\

\end{abstract}

\maketitle

\clearpage
Superconductivity is a strong platform for fault-tolerant quantum computing, provided that long coherence is enabled by the qubit environment. A key challenge in understanding this environment is separating out the different physical mechanisms involved, which can be either classical or quantum in nature. Moreover, the indispensable qubit control can introduce additional sources of decoherence. 

Traditionally, the environment is probed by either relaxation or excitation. Fixed-frequency relaxation measurements have revealed temporal fluctuations \cite{relaxationFixedF_burnett2019decoherence} and spatial correlations \cite{quasiparticle_iaia2022phonon}. Furthermore, spectroscopic measurements of relaxation rates have been applied to identify loss channels \cite{RelaxationSpec_yan2016flux,RelaxationSpec_nguyen2019high,RelaxationSpec_sun2023characterization}, and to uncover the existence of strongly coupled two-level-systems (TLSs) \cite{tlsbarends2013coherent,lisenfeld2015observation,lisenfeld2016decoherence,tlsklimov2018fluctuations, tlslisenfeld2019electric,tlsweeden2025statistics}, thereby guiding improvements in qubit design and materials \cite{surfaceLossImpro_chang2013improved,surfaceLossImpro_place2021new,surfaceLossImpro_bal2024systematic,surfaceLossImpro_biznarova2024mitigation}. And recently, TLSs have been observed to respond hysteretically to electric fields \cite{agarwal2026long}.

In contrast, the excitation process remains underexplored. Excess steady-state excitations are commonly attributed to non-equilibrium quasiparticles \cite{excitation_wenner2013excitation,TempExcitation_jin2015thermal,excitation_serniak2018hot}. Recent observations of long-lived TLSs suggest an additional mechanism: energy stored in these TLSs can flow back into the qubit and re-excite it during quantum algorithms \cite{lltlsspiecker2023two, lltlsgosling2026ll, lltlszhuang2026non}. However, existing studies are restricted to narrowband measurements, spectrally blurring contributions from different noise channels. Reconstructing the full qubit environment requires broadband excitation spectroscopy to separate out the different mechanisms and enable a complete analysis of decoherence. 

Here, we present time-resolved broadband excitation spectroscopy in a frequency-tunable transmon qubit using both postselection and feed-forward. Postselection avoids the reset overhead, making broadband excitation spectroscopy a fast subsequent measurement. With this technique, we reveal the spectroscopic landscape of the excited-state population and qubit effective temperature: a structured background that is interspersed with cold spots, signatures of strongly coupled TLSs. From the excitation and relaxation rates ($\Gup$ and $\Gdown$), we reconstruct the full qubit noise power spectral density (PSD) \cite{Clerknoise2010,vool2017introduction}, which allows us to decompose the background into quantum and classical noise from intrinsic and extrinsic sources. Fitting the quantum noise, we find its effective temperature well above the mixing chamber temperature. This may explain observations of excess excitation in other experiments \cite{excitation_wenner2013excitation,excitation_serniak2018hot,TempExcitation_jin2015thermal, TempExcitation_kulikov2020measuring, TempExcitation_lvov2025thermometry}. 

We also reveal the non-Markovianity of the environment using feed-forward. This technique allows us to enhance the visibility of the long-lived TLSs and acquire broadband statistics about their counts and the environmental lifetime $\tenv$. We observe that long-lived TLSs exist over a wide frequency range. Their frequency distribution and density are consistent with the standard tunneling model (STM). Previously, they have only been reported in fluxonium qubits, where they were associated with JJ arrays  \cite{lltlsspiecker2023two, lltlszhuang2026non,lltlsgosling2026ll}. Our findings instead suggest that they are intrinsic. Model simulations show the existence of three types of dynamics for TLSs within this context: Markovian, resolved non-Markovian, and unresolved non-Markovian. The presence in either regime is primarily defined by the balance between the qubit-TLS coupling $g_{\rm{q-TLS}}$ and the TLS relaxation time $T_{\rm{1,TLS}}$. The model also indicates that, paradoxically, improving $T_1$ suppresses ordinary Markovian relaxation errors but simultaneously exposes environmental memory that was previously hidden by fast qubit decay. Interactions with long-lived TLSs cause time-dependent excitation errors that persist for hundreds of gate cycles. Routine excitation spectroscopy is essential to identify and avoid these, thus bypassing a hidden roadblock for fault-tolerant quantum computation. The presented approach is general and applicable to any frequency-tunable qubit.

\begin{figure}
\centering
\includegraphics{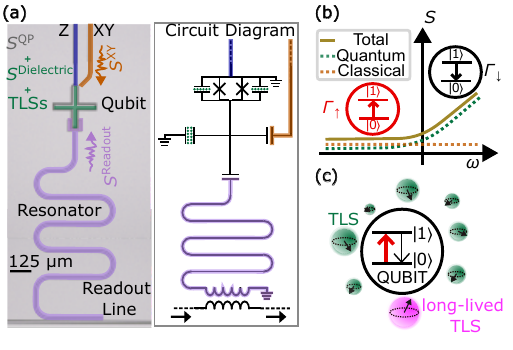} 
\caption{(a) An optical-microscope image of the flux-tunable transmon together with its equivalent circuit. The qubit environment contains extrinsic noise from the drive line (orange) and the readout line (purple). Internal noise originates from TLSs (green) and quasiparticles  (gray). (b) Schematic of the total noise PSD (solid olive line), decomposed into classical noise (dashed orange line) and quantum noise (dashed green line). Negative (positive) frequency corresponds to the excitation (relaxation) process, characterized by the rate $\Gup$ ($\Gdown$). The excitation process used to disentangle these noise sources is indicated in red. (c) Schematic of a fuzzy TLS environment. The long-lived TLS is highlighted in magenta and the excitation due to the TLS is indicated by a red arrow. 
\label{fig:1}}
\end{figure}

Fig.~\ref{fig:1}(a) shows the flux‑tunable transmon qubit (optical microscope image) and a color‑coded map of its surrounding environment and noise sources. The qubit couples not only to external noise from the control and readout lines, but also to quasiparticles and TLSs residing in the qubit environment. Fig.~\ref{fig:1}(b) represents the corresponding noise PSD, which can exhibit either classical or quantum behavior depending on the effective temperature of the noise source \cite{schoelkopf2003qubits, Clerknoise2010}. Here, negative (positive) frequency corresponds to the excitation (relaxation) process, characterized by the rate $\Gup$ ($\Gdown$). In Fig.~\ref{fig:1}(c), we schematically represent the intrinsic qubit materials hosting a large ensemble of TLSs and highlight the presence of long-lived ones, whose existence adds further complexity to the environment.

To precisely separate the classical and quantum noise, broadband excitation spectroscopy is required to characterize the $\Gup$ and $\Gdown$. Relying only on relaxation spectroscopy neglects half of the qubit-environment dynamics. In order to characterize the non-Markovian behavior associated with the long-lived TLSs, the following challenges need to be overcome. The coupling $g_{\rm{q-TLS}}$ needs to be sufficiently strong to populate the TLSs, making them spectroscopically visible. At the same time, the wait time in the protocol needs to be short enough to allow the TLS to be probed before it decays. The former challenge limits the resolution of the method, and the latter is remedied by using feed-forward. In the following, we describe the use of broadband excitation spectroscopy to characterize the excitation process and disentangle different noise aspects.

\begin{figure}
\includegraphics[width=3.37in]{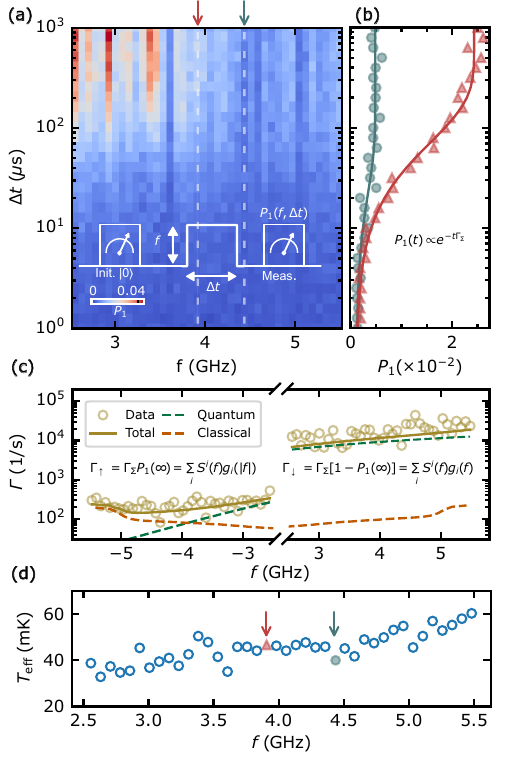}
  \caption{(a) Measured $P_1$ as a function of f and $\Delta t$ after applying excitation spectroscopy. The inset shows the pulse sequence. (b) Example 1d slices for the hot region (red triangles) and the cold region (teal circles). (c) Extracted $\Gup$ and $\Gdown$ as a function of f (yellow circles). Here $\Gup$ corresponds to the negative frequency, while $\Gdown$ corresponds to the positive frequency. The modeled $\Gamma$ from classical noise is shown as a red dashed line. The fitted $\Gamma$ from quantum noise is plotted as a green dashed line. (d) Effective temperature as a function of qubit frequency. 
\label{fig:2}}
\end{figure}

First, we use the technique combined with postselection to differentiate between quantum and classical noise. The postselection eliminates the necessity of reset, allowing fast broadband excitation spectroscopy. The results in Fig.~\ref{fig:2}(a) show a clear time-resolved spectroscopic landscape of excited-state population $P_1$, incorporating a structured background interspersed with cold regions that correspond to the presence of strongly coupled TLSs. The pulse scheme is shown in the inset. Two data slice examples and their fitted lines ($P_1$ versus $\Delta t$) for a hot and a cold region are shown in Fig.~\ref{fig:2}(b). At each frequency, we fit the slices with
\begin{equation}
P_1(t) = [P_1(0) - P_1(\infty)] \exp(-(\Gamma_\uparrow+\Gamma_\downarrow)t) + P_1(\infty), 
\label{eq:excitation}
\end{equation} 
where \(P_1(\infty) = \Gamma_\uparrow/(\Gamma_\uparrow+\Gamma_\downarrow)\). The fit yields $\Gup(f)$ and $\Gdown(f)$, which are plotted in Fig.~\ref{fig:2}(c). Exploiting the detailed balance relation, we extract the qubit effective temperature $T_{\rm{eff}}$ as a function of qubit frequency from the transition rates, and the result is shown in Fig.~\ref{fig:2}(d).

The non-monotonic behavior of the background excitation indicates that the qubit couples to multiple noise sources. We use the extracted $\Gamma (f)$ to reconstruct the environment by modeling the extrinsic noise from the microwave components, and by fitting for the effective temperature of the intrinsic noise (see Sec.~IV and V of the Supplemental Material~\cite{supp}). This reconstruction enables us to decompose the noise into classical and quantum contributions. We validate the classical noise model by comparing the modeled curve with the data. Assuming the remaining noise originates from a single thermal bath, we extract an effective temperature of 38~mK, consistent with quantum noise. Although excess $\Gup$ is commonly attributed to non-equilibrium quasiparticles, our result shows that quantum noise or microwave-component-induced classical noise may account for the observed excitation in other experiments \cite{excitation_wenner2013excitation,excitation_serniak2018hot,TempExcitation_jin2015thermal, TempExcitation_kulikov2020measuring, TempExcitation_lvov2025thermometry}. Excess excitation arising from classical noise has long been recognized in the field. However, we present the first direct quantitative observation of its impact on excitation. We also show that the qubit can act as a sensitive in-situ spectrometer of noise originating from the control electronics (see Sec.~IV of the Supplemental Material~\cite{supp}). We find the qubit effective temperature $T_{\rm{eff}}$ is not a single fixed value, but depends on transition frequencies, varying by up to 50$\%$ across the measured band. Our method provides a direct probe of the excitation process, whereas traditional relaxation spectroscopy inherently excites long-lived TLSs, biasing the characterization of qubit excitation by the non-Markovian behavior (see Sec.~III of the Supplemental Material~\cite{supp}).

\begin{figure}
\includegraphics{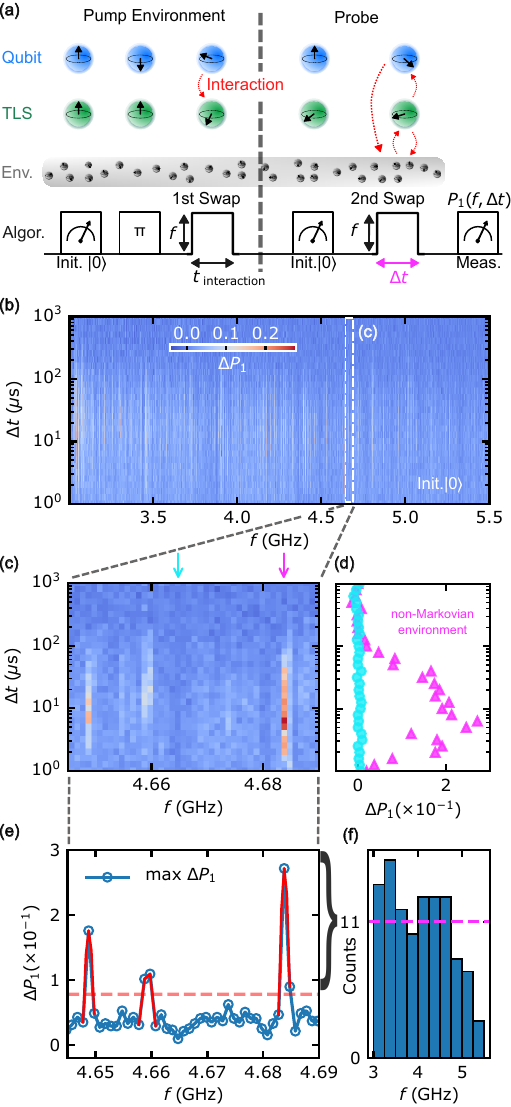}
  \caption{(a) Pulse sequence used to separate Markovian and non-Markovian environments. It consists of an environmental‑pump operation followed by a probe operation using excitation spectroscopy. The red dotted arrows signify the interaction between the qubit, the TLS and the environment. (b) Measured results are shown as $\Delta P_1$ versus f and $\Delta t$. Here, $\Delta P_1$ denotes the measured $P_1$ after subtraction of the thermal excitation background. (c) A zoomed-in plot of $\Delta P_1$ versus f and $\Delta t$ is shown. (d) Example 1d slices of $\Delta P_1$ versus $\Delta t$ for Markovian (cyan circles) and non-Markovian (magenta triangles) environments. (e) The maximum $\Delta P_1$ for each frequency is plotted as blue circles. The extracted peaks are indicated by red lines. The threshold, $1 \sigma$ above the mean, is indicated by the red dashed line. (f) The histogram for the non-Markovian TLSs uses a 250~MHz bin size. The expected counts are shown as a magenta dashed line.
\label{fig:3}}
\end{figure}

A populated non-Markovian environment causes additional qubit excitation. To properly characterize the non-Markovian environment and separate it from the Markovian environment, we combine the environmental-pump operation with a probe operation using excitation spectroscopy, see Fig.~\ref{fig:3}(a). In the first stage, we DC bias the qubit at the sweet spot, initialize the qubit in the $\ket{0}$ state using feed-forward, drive the qubit to the $\ket{1}$ state with a $\pi$ pulse, and excite the TLSs by flux-tuning the qubit to various frequencies $f$ for an interaction time chosen to be $t_{\rm{interaction}} = 10~\mu s$. Next, we perform the excitation spectroscopy, similar to the one shown in Fig.~\ref{fig:2}(a), to probe the environmental excitation. 

At each frequency point, in addition to measurements with the pump sequence, we perform reference measurements without the pump sequence to characterize the background thermal excitation. The resulting time- and frequency-dependent excess population, $\Delta P_1$, after subtracting the thermal contribution, is shown in Fig.~\ref{fig:3}(b,c). The result shows excess qubit excitation clustered between 1 and 100 $\mu s$ over the entire frequency range. Two representative $\Delta P_1$ versus $\Delta t$ slices for Markovian (cyan circles) and non-Markovian (magenta triangles) environments are shown in Fig.~\ref{fig:3}(d). For a non-Markovian environment, the excess qubit excitation $\Delta P_1$ first increases, reaches a maximum, and then decreases. The initial $\Delta P_1$ rise is caused by energy flowing back to the qubit from the populated environment. During this process, the energy continuously decays from both the qubit and the long-lived TLSs, leading to an eventual reduction in $\Delta P_1$. In contrast, a Markovian environment yields $\Delta P_1 \approx 0$. We also extract the maximum $\Delta P_1$ at each frequency and use a peak finding algorithm to identify long-lived TLSs, see Fig.~\ref{fig:3}(e). The threshold is $1 \sigma$ above the mean as indicated by the red dashed line. We then count the number of long-lived TLSs and the histogram constructed with a 250~MHz bin size is shown in Fig.~\ref{fig:3}(f).

The resolved long-lived TLSs are uniformly distributed across the 3-4.5~GHz frequency range, in agreement with the STM prediction \cite{enss2005low}
    \begin{equation}
        D(f,\ttls) = \frac{D_0}{2\ttls \sqrt{1-T_{\rm{1,TLS}}^{\rm{min}}(f)/T_{\rm{1,TLS}}}}
        \label{eq:tls_distribution}.
    \end{equation}
 In the long-lived TLS limit, $T_{\rm{1,TLS}} \gg T_{\rm{1,TLS}}^{\rm{min}}(f)$, and the distribution becomes uniform. Here $D$ is the TLS density distribution and $D_0$ is a constant. Above 4.5~GHz, the counts are noticeably reduced. We interpret this reduction as a consequence of increased qubit decay rate in that spectral region. The measured density of the long-lived TLSs is 44~counts/GHz. To connect the observed density with the STM, we use the effective volume of the dielectric that is 2~nm thick, 2000~$\mu m$ long (perimeter of the qubit capacitor), and 50~nm wide based on \cite{bilmes2020resolving, bilmes2022probing}. We obtain $D_0 \sim 1.5 \times 10^{44} J^{-1} m^{-3}$, corresponding to a surface density of $0.2~\rm{GHz}^{-1} \mu \rm{m}^{-2}$, in line with the previous report on fluxonium qubits \cite{lltlszhuang2026non}. From the extracted surface density, we plot the expected counts in Fig.~\ref{fig:3}(f). The long-lived TLSs have a distribution and density that agree with the STM and other reports, indicating that long-lived TLSs are not specific to fluxonium qubits with a junction chain, but rather a general feature of the qubit environment.

The density of the long-lived TLSs in our device is sufficiently high to affect the qubit operations. Fig.~\ref{fig:3}(e) shows that each long-lived TLS has a linewidth of more than 1~MHz. On average, each long-lived TLS occupies a bandwidth of $\sim$2~MHz. As a result, about $9 \%$ of the qubit-frequency band is affected by these defects. In addition, the frequency of the TLSs is known to drift slowly due to the fluctuations in the electric field and local strain \cite{tlsklimov2018fluctuations}, further shrinking the usable spectral window. To guarantee reliable operation, routine excitation spectroscopy is required to identify and bypass long-lived TLSs.

\begin{figure}
\centering
\includegraphics[width=3.37in]{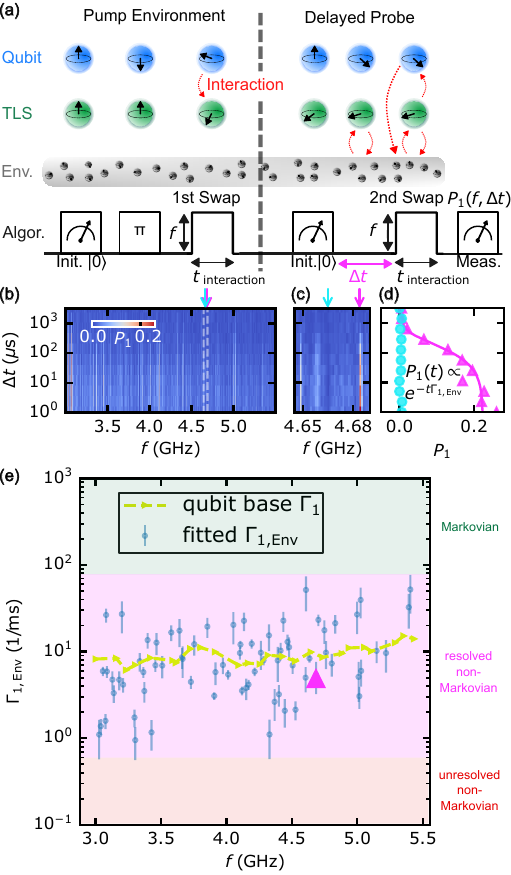}
  \caption{(a) Pulse sequence used to extract $T_{\rm{1,Env}}$. The red dotted arrows signify the interaction among the qubit, the TLS and the environment. (b) Measured results are shown as $P_1$ versus f and $\Delta t$. (c) A zoomed-in plot of $P_1$ versus f and $\Delta t$. (d) Example 1d slices of $P_1$ versus $\Delta t$ for Markovian (cyan circles) and non-Markovian (magenta triangles) environments. (e) The measured $\Gamma_{\rm{1, Env}}$ versus f is shown as the blue circles with error bars obtained from the fit. The yellow dashed line represents the qubit baseline $\Gamma_1$. The Markovian regime is defined by an environmental relaxation rate sufficiently large that the excitation decays before being re-excited by the qubit, precluding data points in this regime. Conversely, in the unresolved non-Markovian regime, the environment has a longer lifetime than the qubit, but the coupling is too weak to excite it, thereby precluding a measurable signal.
\label{fig:4}}
\end{figure}

With the long-lived TLSs identified and the statistics computed, we can now quantify the memory time $\tenv$ of the environment. The pulse sequence in Fig.~\ref{fig:4}(a) is a slight modification of that shown in Fig.~\ref{fig:3}(a): we vary the delay times $\Delta t$ before the second swap while keeping $t_{\rm{interaction}} = 10~\mu s$. The result ($P_1$ versus $f$ and $\Delta t$), which shows the decay of the environmental excitation, is shown in Fig.~\ref{fig:4}(b,c). Two representative $P_1$ versus $\Delta t$ slices for Markovian (cyan circles) and non-Markovian (magenta triangles) environments are shown in Fig.~\ref{fig:4}(d). For a non-Markovian environment, the excess qubit excitation $P_1$ decays to the phonon bath as the delay time $\Delta t$ increases. From the fits, we extract the environmental lifetime $\tenv$ and plot the corresponding decay rate $\Gamma_{\rm{1, Env}} = 1/T_{1, \rm{Env}}$ along with the background qubit decay rate, see Fig.~\ref{fig:4}(e). Here, we show the non-Markovian environment caused by the long-lived TLSs. Excluding data with a fitting error exceeding $ 50 \%$, we find that 58$\%$ of the long-lived TLSs have lifetimes longer than that of the qubit.

The distribution of $\Gamma_{\rm{1,Env}}$ is uniform across frequencies, as predicted by Eq.~\ref{eq:tls_distribution}. Most of the $\Gamma_{\rm{1, Env}}$ values are clustered between 1 ms$^{-1}$ and 50 ms$^{-1}$. At the high $\Gamma_{\rm{1, Env}}$ end, the reduced visibility is explained by strong coupling to the phonon bath \cite{enss2005low, yu2004study}, where a large fraction of TLS excitation decays to the phonon bath instead of re-exciting the qubit. At low $\Gamma_{\rm{1, Env}}$, the reduced counts cannot be explained by this mechanism. Instead, it reflects an intrinsic link between the qubit-TLS coupling 
 \begin{equation}
        \gqtls = \frac{2}{\hbar}\,(\mathbf{d} \cdot \mathbf{E})\,\frac{\Delta_0}{f}
        \label{eq:gqtls}
\end{equation}
and the TLS relaxation rate
\begin{equation}
    \Gamma_{\rm{1, TLS}} = A \Delta_0^2 f, \qquad A = \left(
    \frac{1}{v_\ell^5}
    + \frac{2}{v_t^5}
    \right) \frac{\gamma^2}{(16\pi^4 \rho \hbar)}
        \label{eq:gamma_tl}
\end{equation} 
for $ k_B T \ll hf$ \cite{yu2004study,enss2005low,Muller2019TLS}. Here, $\mathbf{d}$ is the electric dipole moment of the TLS, $\mathbf{E}$ the zero-point electric field of the qubit, $\Delta_0$ the TLS tunneling frequency, $f$ the TLS frequency, $\gamma$ the deformation potential energy, $\rho$ the mass density, the $v_{\ell}$ and $v_{t}$ are longitudinal and transverse speeds of sound. Both quantities depend on the tunneling frequency $\Delta_0$. As we can see from the above equations, a low tunneling frequency $\Delta_0$ yields a slow TLS decay $\Gamma_{\rm{1, TLS}}$ and a small qubit-TLS coupling $\gqtls$. In this regime, we cannot excite the TLS with slow decay ($\Gamma_{\rm{1, TLS}} < $ 1 ms$^{-1}$) efficiently, and those TLSs are effectively invisible to our measurement. Our model reproduces the suppressed counts at $\Gamma_{\rm{1, TLS}} \approx 1$ ms$^{-1}$ (see Sec.~VIII of the Supplemental Material~\cite{supp}). Paradoxically, this model predicts that as the qubit $\tq$ improves, previously unresolved long-lived TLSs can become visible, and give rise to non-Markovian errors that would otherwise go unnoticed. As such, these findings are distinct from the notion underlying continuous qubit calibration that the qubit environment is Markovian \cite{tlsklimov2018fluctuations}. Excitation spectroscopy can expose this otherwise hidden roadblock and maintain qubit performance for fault-tolerant quantum computing.
 
In summary, we introduce broadband excitation spectroscopy to characterize the qubit environment and apply it to a frequency-tunable transmon. We find that the qubit excitation exhibits a varying landscape interleaved with cold spots. We extract the relaxation ($\Gdown$) and excitation ($\Gup$) rates to disentangle classical and quantum noise arising from intrinsic and extrinsic sources. Instrumental classical noise or elevated temperature of the quantum noise provides a plausible explanation for the excess qubit excitation observed in other reports. We also resolve long-lived TLSs in a wide frequency range and find that their statistics match the predictions of the STM. Our results support long-lived TLSs as a generic feature of superconducting qubit environments rather than a peculiarity of the fluxonium junction arrays. Our model predicts that, paradoxically, as the qubit $T_1$ improves, the qubit re-excitation process intensifies. The decay rates of the non-Markovian environment are roughly 1-50~ms$^{-1}$, corresponding to a memory time of 20~$\mu s$ to 1~ms. For a typical QEC cycle time of $\approx$1~$\mu$s \cite{GoogleQuantumAI2023SurfaceCode}, the non-Markovian environment can retain memory for hundreds of QEC cycles. The environment produces history-dependent, non-Markovian qubit errors that lead to protocol-dependent gate performance \cite{gate_yan2018tunable,gate_nakamura2024gate,gate_gao2025ultrafast,gate_nakamura2026entanglement}. The presented approach separates different noise mechanisms, quantifies the excitation process and long-lived TLS densities, and helps pave the way to stable qubit performance (memory) for fault-tolerant quantum computation.

\section{acknowledgement}
We thank R. Hanna for valuable discussions, M. Guardascione for assistance with the TWPA, and N. Oertel for support in the automated four‑point‑probe measurement system. The device was fabricated at the Helmholtz Nano Facility (HNF) at Forschungszentrum Jülich. We thank the German Federal Ministry of Research, Technology and Space (BMFTR) for financial support through the “Quantum Technologies – from Basic Research to Market” programme (project QSolid, Grant No. 13N16149).

\section{contributions}
Y.L., J.R.G., and R.B. designed the experiments. J.R.G. and Y.L. executed the measurements. Y.L. and J.R.G. performed the data analysis. The data‑acquisition software was developed by J.R.G., Y.L., Y.G., A.G., J.T.S., and J.C. The device was designed by Y.G.. Fabrication was carried out jointly by Y.H., D.A.V., J.R.G., H.B., M.N., and Y.L.. The manuscript was written by Y.L., J.R.G., and R.B. All authors contributed to the underlying experimental infrastructure, the theoretical interpretation, and the critical revision of the manuscript.

\nocite{lambert2026qutip,gustavsson2016suppressing}

\bibliography{references}

\end{document}


\title{Supplementary Materials for “Quantum environment afterglow from broadband excitation spectroscopy in superconducting qubits”}
\author{J. R. Guimarães}
\email[]{j.guimaraes@fz-juelich.de}
\thanks{These authors contributed equally}
\affiliation{Institute for Functional Quantum System (PGI-13), Forschungszentrum Jülich, 52425 Jülich, Germany}
\affiliation{Department of Physics, RWTH Aachen University, 52074 Aachen, Germany}

\author{Y.~Liu}
\email[]{ye.liu@fz-juelich.de}
\thanks{These authors contributed equally}
\affiliation{Institute for Functional Quantum System (PGI-13), Forschungszentrum Jülich, 52425 Jülich, Germany}

\author{Y.~Gao}
\affiliation{Institute for Functional Quantum System (PGI-13), Forschungszentrum Jülich, 52425 Jülich, Germany}
\affiliation{Department of Physics, RWTH Aachen University, 52074 Aachen, Germany}

\author{Y.~Haddad}
\affiliation{Institute for Functional Quantum System (PGI-13), Forschungszentrum Jülich, 52425 Jülich, Germany}
\affiliation{Department of Physics, RWTH Aachen University, 52074 Aachen, Germany}

\author{A.~Galicia}
\affiliation{Institute for Functional Quantum System (PGI-13), Forschungszentrum Jülich, 52425 Jülich, Germany}
\affiliation{Department of Physics, RWTH Aachen University, 52074 Aachen, Germany}

\author{D.~A.~Volkov}
\affiliation{Institute for Functional Quantum System (PGI-13), Forschungszentrum Jülich, 52425 Jülich, Germany}
\affiliation{Department of Physics, RWTH Aachen University, 52074 Aachen, Germany}

\author{J.~T.~Schmieder}
\affiliation{Institute for Functional Quantum System (PGI-13), Forschungszentrum Jülich, 52425 Jülich, Germany}
\affiliation{Department of Physics, RWTH Aachen University, 52074 Aachen, Germany}

\author{H.~Bhardwaj}
\affiliation{Institute for Functional Quantum System (PGI-13), Forschungszentrum Jülich, 52425 Jülich, Germany}
\affiliation{Department of Physics, RWTH Aachen University, 52074 Aachen, Germany}

\author{M.~Neis}
\affiliation{Institute for Functional Quantum System (PGI-13), Forschungszentrum Jülich, 52425 Jülich, Germany}
\affiliation{Department of Physics, RWTH Aachen University, 52074 Aachen, Germany}

\author{J.~Cereijo}
\affiliation{Institute for Functional Quantum System (PGI-13), Forschungszentrum Jülich, 52425 Jülich, Germany}
\affiliation{Department of Physics, RWTH Aachen University, 52074 Aachen, Germany}

\author{M.~Jerger}
\affiliation{Institute for Functional Quantum System (PGI-13), Forschungszentrum Jülich, 52425 Jülich, Germany}

\author{P.~A.~Bushev}
\affiliation{Institute for Functional Quantum System (PGI-13), Forschungszentrum Jülich, 52425 Jülich, Germany}

\author{R.~Barends}
\email[]{r.barends@fz-juelich.de}
\affiliation{Institute for Functional Quantum System (PGI-13), Forschungszentrum Jülich, 52425 Jülich, Germany}
\affiliation{Department of Physics, RWTH Aachen University, 52074 Aachen, Germany}

\date{\today}
\maketitle

\setcounter{secnumdepth}{3}   
\renewcommand\thesection{\Roman{section}}
\setcounter{figure}{0} 
\setcounter{equation}{0}
\setcounter{table}{0}
\renewcommand{\thefigure}{S\arabic{figure}}  
\renewcommand{\theequation}{S\arabic{equation}}
\renewcommand{\thetable}{S\arabic{table}}
\renewcommand{\thesection}{\Roman{section}}

\section{Device fabrication}
\label{sec:fab}

We use a high-resistivity intrinsic silicon wafer ($\rho >$ 10~k$\Omega \cdot {\rm cm}$). It is cleaned sequentially in acetone, isopropanol (IPA) and de-ionized (DI) water, followed by a 5-minute O$_2$ plasma clean to remove organic contamination. A bilayer resist of SF5 and UV6 is spin-coated. Subsequently, the wafer is loaded into the electron beam lithography system to pattern both the large structures (ground plane, capacitors, resonators), and small structures such as 100~nm wide Josephson junctions.

The wafer is developed in MIF-326 developer, followed by a mild O$_2$ plasma clean to remove the resist residue. A quick 1$\%$ HF dip removes the native oxide on the exposed Si, after which the wafer is rinsed with water until the water resistivity is above 10~M$\Omega$. The wafer is transferred into the ebeam evaporator within 5 minutes after the rinsing to prevent oxide regrowth. The ebeam evaporator has a separate chamber for oxidation. After standard double angle evaporation for the Manhattan-style Josephson junction, we lift off the wafer in two consecutive DMSO baths at 80~$^\circ$C. Finally, the wafer is diced into 1~cm $\times$ 1~cm chips.

\section{Experimental setup}
\begin{figure}[b]
\centering
\includegraphics[width=3.37in]{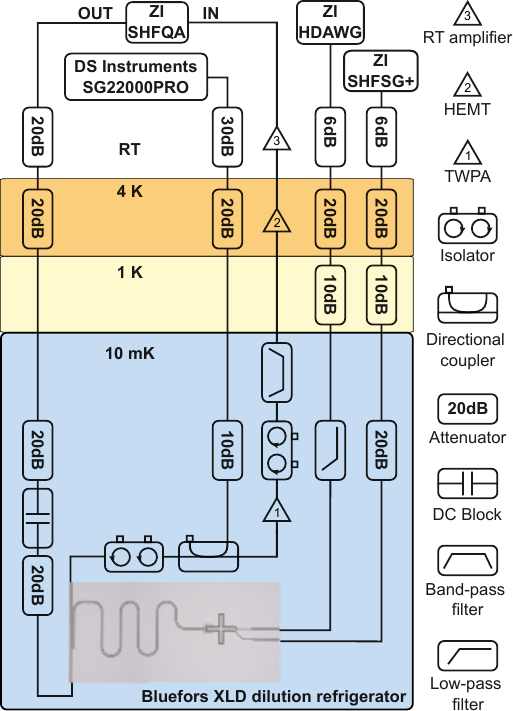}
  \caption{Schematic of electronic devices and dilution fridge wiring.
\label{figs:FigS_wiring}}
\end{figure}

The device is packaged in a light-tight sample box and encapsulated by a magnetic shield. The whole sample stage is mounted in a dilution fridge with a base temperature of $10$~mK. The wiring diagram of the dilution fridge is displayed in Fig.~\ref{figs:FigS_wiring}. Our cryogenic setup is similar to the configuration used in our prior work \cite{gao2025ultrafast}. A ZI SHFSG+ generates the drive microwave tone while the ZI SHFQA generates and detects the readout tone. The qubit frequency is tuned with a ZI HDAWG. These lines are strongly attenuated and filtered at multiple stages to suppress the noise from the machine and from higher temperature stages of the fridge. A travelling-wave parametric amplifier (TWPA) increases the signal-to-noise ratio, making single-shot readout possible. Flux predistortion on the ZI HDAWG is applied to accurately control the qubit frequency and reduce flux distortion. 

\section{Relaxation-based method and long-lived TLS}

\begin{figure}
\centering
\includegraphics[width=3.37in]{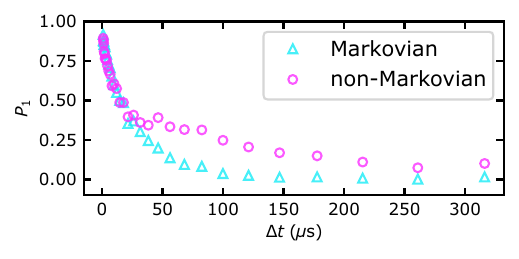}
  \caption{Qubit relaxation dynamics when interacting with Markovian and non-Markovian environments. 
\label{figS:plot_p1_vs_time}}
\end{figure}
When the qubit interacts with a Markovian environment, the relaxation is purely exponential. In contrast, when the qubit interacts with long-lived TLSs, the energy stored in these long-lived TLSs flows back to the qubit, producing a non-exponential decay as shown in Fig.~\ref{figS:plot_p1_vs_time}. Using relaxation spectroscopy to identify long-lived TLSs relies on whether the process strictly follows an exponential decay. However, it is not reliable because 1) the deviation may be too subtle, making the method insensitive to weakly coupled long-lived TLSs, 2) other mechanisms, such as quasiparticle tunneling \cite{gustavsson2016suppressing} or TLS frequency drift, can also produce similar non-exponential decay, leading to misidentification of long-lived TLSs.

\section{Noise modeling from drive line}

\begin{figure}
\centering
\includegraphics[width=3.37in]{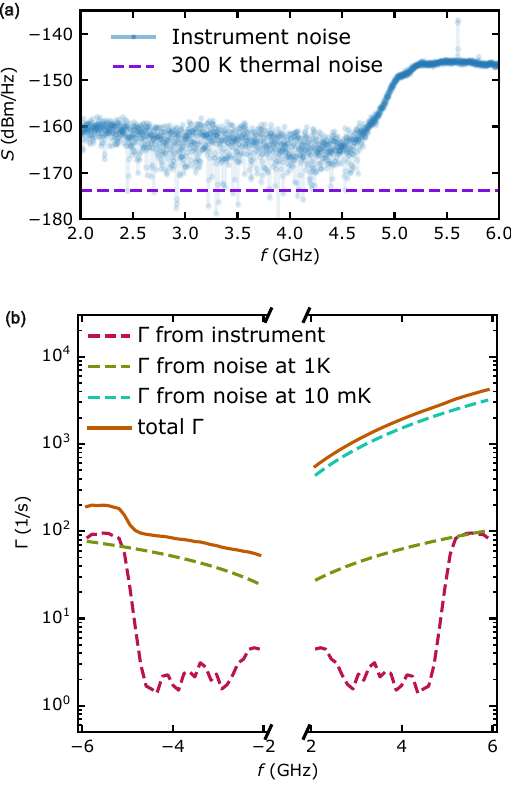}
  \caption{(a) Measured instrument noise. (b) Calculated transition rate $\Gamma$ from instrument noise and classical noise originating from components at effective temperature of 1~K and 10~mK.
\label{figs:classical_noise}}
\end{figure}

To model the classical noise from the drive line, we first determine the coupling capacitance $C_c$ between the drive line and the qubit capacitor. We measure the required instrument voltage amplitude $A_0$ for a $\pi$ pulse by performing a Rabi experiment. The signal is attenuated by the attenuators and cables before reaching the chip. We use a cosine drive pulse. The coupling capacitance is extracted from 
\begin{equation}
    C_c = \frac{1}{A t} \sqrt{\frac{{\pi \hbar} C}{f}},
\end{equation} where $C$ is the total qubit capacitance, $f$ is the qubit frequency, $A$ is the on-chip Rabi amplitude estimated from the known attenuation and cable loss, and $t$ is the duration of the Rabi drive.

Now, we calculate the transition rate estimated using Fermi's golden rule
\begin{equation}
    \Gamma = \frac{K^2}{\hbar^2} S(f)
    \label{eq:FermisGoldenRule},
\end{equation}
where K is the coupling constant \cite{Clerknoise2010}. For a capacitively coupled drive line, 
\begin{equation}
K^2 = \frac{ \hbar \pi \abs{f} C_c^2}{C}.
\end{equation}

The noise PSD $S(f)$ can arise from an active source (the control instrument), or from finite-temperature passive elements (attenuators and cables). 

To model the noise PSD, we first measure the output noise of the drive-control instrument ZI-SHFSG+ with the band‑pass filter set to 5–10~GHz. The background of the spectrum analyzer was calibrated by terminating its input with a 50~$\Omega$ load and recording the background level $N_0$. Connecting the ZI-SHFSG+ to the spectrum analyzer gave a total input noise $N_1$. The instrument noise is obtained as $N_1 - N_0$ and is shown in Fig.~\ref{figs:classical_noise}(a). The noise remains low up to 4.6~GHz and then jumps to -146~dBm/Hz above that frequency. 

The signal then passes through 56 dB attenuation and cable loss. The attenuators and cables reside at finite temperatures (50~K, 4~K, 1~K, 10~mK) and emit thermal noise 
\begin{equation}
S_V(\omega)
=
\begin{cases}
2\hbar\omega R \left[n_B(\omega)+1\right]
& \omega>0, \\[6pt]
2\hbar|\omega| R\, n_B(|\omega|)
& \omega<0,
\end{cases}
\end{equation} where $n_B(\Omega)=\frac{1}{e^{\hbar\Omega/k_B T}-1}$. An attenuator with attenuation factor $\alpha$ passes only the fraction (1-$\alpha$) of its own noise to the colder stage. The cumulative noise arriving at the drive line of the qubit is therefore the sum of the attenuated contributions from all temperature stages. 

We calculate the $\Gamma$ contributions from individual components by applying Eq.~\ref{eq:FermisGoldenRule}, as shown in Fig.~\ref{figs:classical_noise}(b). The positive frequencies correspond to the relaxation rate $\Gdown$ and the negative frequencies to the excitation rate $\Gup$. The relaxation rate is almost entirely determined by the quantum noise, whereas the excitation rate is dominated by the classical noise from the instrument and from the 1~K stage. The direct observation of the classical-noise-induced excitation, although expected, has not been quantified before. The modeled excitation rate $\Gup$ reproduces the experimental trend within measurement uncertainty.

\section{Fitting quantum noise}
\label{sec:fit_quantum_noise}
Having modeled both the classical and quantum noise contributions from various components, we now model the remaining intrinsic noise as originating from a single thermal bath associated with the intrinsic materials. The relaxation rate is described by $\Gdown = \omega /Q$. The fitted Q is 1.6~M. Subsequently, we fit the ratio between $\Gup$ and $\Gdown$ with the detailed-balance relation \cite{Clerknoise2010}:
\begin{equation}
    \frac{\Gup}{\Gdown} = \exp{-\frac{\hbar \omega}{k_{\rm{B}} T}}.
\end{equation}
The resulting fitted temperature is 38~mK.

\section{qubit-TLS interaction}

\begin{figure}
\centering
\includegraphics[width=3.37in]{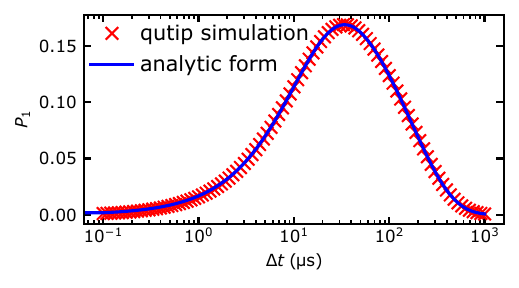}
  \caption{Time evolution of the qubit excited‑state population during interaction with a long‑lived TLS obtained from QuTiP simulations and compared with the corresponding analytical solution.
\label{figs:p1_qutip_simulation}}
\end{figure}

We model the qubit-TLS system with the Lindblad master equation
\begin{equation*}
\dot{\rho}
=
-i[H,\rho]
+
\sum_k
\mathcal D[L_k]\rho,
\qquad
\mathcal D[L]\rho
=
L\rho L^\dagger
-
\frac12
\left\{
L^\dagger L,\rho
\right\},
\end{equation*}
where the Lindblad terms $L_i$ are 
\begin{equation*}
\sqrt{\Gamma_{\rm 1,q}}\,\sigma_-^{(q)},
\sqrt{\Gamma_{\rm \phi, q}/2}\,\sigma_z^{(q)},
\sqrt{\Gamma_{\rm 1, TLS}}\,\sigma_-^{(t)},
\sqrt{\Gamma_{\rm \phi, TLS}/2}\,\sigma_z^{(t)},
\end{equation*} with i = 1,2,3,4. Here, $\Gamma_{\rm 1,q}$ is the qubit relaxation rate, $\Gamma_{\rm \phi, q}$ is the qubit pure dephasing rate, $\Gamma_{\rm 1, TLS}$ is the TLS relaxation rate, $\Gamma_{\rm \phi, TLS}$ is the TLS dephasing rate.

From the master equation, we obtain the rate equations
\begin{align}
\dot p_q
&=
-\Gamma_{\rm 1,q}p_q
+
\Gamma_{\rm qt}(p_t-p_q),
\\
\dot p_t
&=
-\Gamma_{\rm 1, TLS}p_t
+
\Gamma_{\rm qt}(p_q-p_t),
\end{align}
with the incoherent exchange rate
\begin{equation}
\Gamma_{\rm qt}=\frac{2g^2\Gamma_c}{\Delta^2+\Gamma_c^2}.
\label{eq:Gamma_ex}
\end{equation}
Here, $\Gamma_c$ is the combined decoherence rate of the qubit-TLS system. 

In our experiment for qubit-TLS interaction, we assume the TLS is populated with $p_t(0) = p_0$, and the qubit is reset to the ground state with $p_q(0)=0$. The solution of Eq.~\ref{eq:Gamma_ex} is 
\begin{equation}
P_q(t)
=
p_0
\frac{\Gamma_{\rm qt}}{\Lambda}
\left(
e^{\lambda_+t}
-
e^{\lambda_-t}
\right),
\label{eq:Pq_solution}
\end{equation}
where
\begin{align}
\Lambda
&=
\sqrt{
(\Gamma_{\rm 1,q}-\Gamma_{\rm 1, TLS})^2
+
4\Gamma_{\rm qt}^2},
\\
\lambda_\pm
&=
-
\frac{
\Gamma_{\rm 1,q}
+
\Gamma_{\rm 1, TLS}
+
2\Gamma_{\rm qt}
}{2}
\pm
\frac{\Lambda}{2}.
\end{align}
This solution holds in the limit $\Gamma_c \gg $g, where coherent oscillations are suppressed and the dynamics reduce to a purely biexponential population exchange.

To validate this solution, we also performed simulations of the qubit-TLS interaction using quantum toolbox in Python (QuTiP) \cite{lambert2026qutip}, and the numerical results agree with the analytic calculation. The simulation result and analytic solution are shown in Fig.~\ref{figs:p1_qutip_simulation} with the parameters: $T_{\rm 1,q} = 100~\mu {\rm s}$, $T_{\rm \phi,q}=10~\mu {\rm s}$, $T_{\rm 1,TLS} = 500~\mu {\rm s}$, $T_{\rm \phi, TLS} = 50~{\rm ns}$, $g=100~{\rm kHz}$.

\section{TLS distribution}

Based on the standard tunneling model (STM), TLSs can be characterized by an asymmetry energy $\Delta$ and a tunneling frequency $\Delta_0$. The transition frequency follows $f=\frac{\sqrt{\Delta^2+\Delta_0^2}}{h}$,
with $h$ being Planck's constant \cite{enss2005low}. The distributions $D(\Delta,\lambda)$ of $\Delta$ and tunneling parameter $\lambda$ are independent and uniform
\begin{equation}
D(\Delta,\lambda)\,d\Delta\,d\lambda
= D_0\,d\Delta\,d\lambda.
\end{equation}
Here $\Delta_0=h\Omega e^{-\lambda}$ and h$\Omega$ is the intrawell energy scale \cite{enss2005low}. 

Performing a Jacobian transformation and changing variables from $(\Delta,\lambda)$ to $(f,\Delta_0)$ gives
\begin{equation}
D(f,\Delta_0)\,d\Delta_0\,df
=
D_0
\frac{hf}
{\Delta_0
\sqrt{(hf)^2-\Delta_0^2}}
\,d\Delta_0\,df.
\label{eq:P_f_Delta0}
\end{equation}

The relaxation rate of a TLS is
\begin{equation}
    \Gamma_{\rm{1, TLS}} = A \Delta_0^2 f, \qquad A = \left(
    \frac{1}{v_\ell^5}
    + \frac{2}{v_t^5}
    \right) \frac{\gamma^2}{(16\pi^4 \rho \hbar)}
        \label{eq:gamma_tl}
\end{equation} 
for $ k_B T \ll hf$ \cite{yu2004study,enss2005low}. Here, $\Delta_0$ is the TLS tunneling frequency, $f$ the TLS frequency, $\gamma$ the deformation potential energy, $\rho$ the mass density, $v_{\ell}$ and $v_{t}$ are the longitudinal and transverse speed of sound, respectively.

Using the relaxation time $T_{\rm 1, TLS} = \frac{1}{\Gamma_{\rm 1, TLS}}$ and applying the same Jacobian transformation to Eq.~\ref{eq:gamma_tl} leads to
\begin{equation}
D(f,T_{\rm 1, TLS})
=
\frac{D_0}
{2T_{\rm 1, TLS}}
,
\label{eq:P_tau}
\end{equation}

Thus, for a given frequency, the relaxation times $T_1$ are broadly distributed, indicating the coexistence of Markovian and non-Markovian TLSs. The presence of non-Markovian TLSs appears to be an intrinsic property of the qubit environments.

\section{Reduced visibility due to connection between coupling and relaxation}

\begin{figure}
\centering
\includegraphics[width=3.37in]{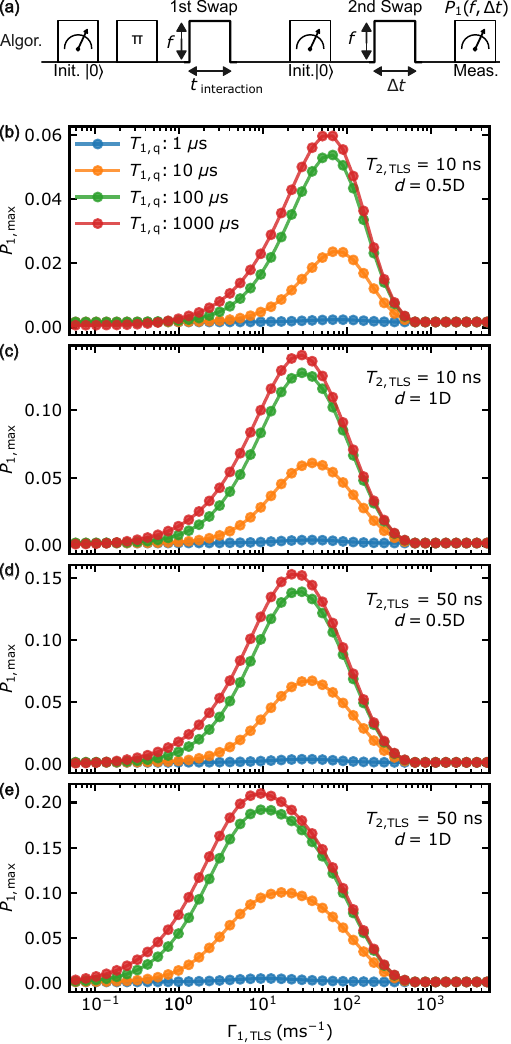}
  \caption{$P_{1,\rm{max}}$ versus $\Gamma_{1,\rm{TLS}}$ with different parameters: (a) $T_{2,\rm{TLS}}=10$~ns and dipole $\mathbf{d}=0.5$~D. (b) $T_{2,\rm{TLS}}=10$~ns and dipole $\mathbf{d}=1$~D. (c) $T_{2,\rm{TLS}}=50$~ns and dipole $\mathbf{d}=0.5$~D. (d) $T_{2,\rm{TLS}}=50$~ns and dipole $\mathbf{d}=1$~D. This indicates the existence of optimal $\Gamma_{1,\mathrm{TLS}}$ windows for detecting long-lived TLSs.
\label{figs:p1_max_vs_gamma_tls}}
\end{figure}

Here, we explain the reduced visibility of the long-lived TLSs with $\ttls >$1~ms by connecting the coupling strength and TLS relaxation. The qubit-TLS coupling is
\begin{equation}
    \gqtls = \frac{2}{\hbar}\,(\mathbf{d} \cdot \mathbf{E})\,\frac{\Delta_0}{f}    
\label{eq:coupling_gqtls},
\end{equation}where $\mathbf{d}$ is the electric dipole moment of the TLS, $\mathbf{E}$ the zero-point electric field associated with the qubit, $\Delta_0$ the TLS tunneling frequency, $f$ the TLS frequency.

The qubit-TLS coupling Eq.~\ref{eq:coupling_gqtls} and TLS relaxation Eq.~\ref{eq:gamma_tl} both depend on $\Delta_0$. As a result, for a given material (fixed deformation potential, mass density, speed of sound and electric dipole moment), and for fixed qubit parameters (electric field and qubit frequency), the relaxation rate and coupling are directly connected. This intrinsic connection suppresses the visibility of long-lived TLSs when their lifetime is longer than 1~ms, because of the weak coupling.

To quantify this effect, we use the QuTiP toolbox, see Fig.~\ref{figs:p1_max_vs_gamma_tls}. In Fig.~\ref{figs:p1_max_vs_gamma_tls}(a), we show the simulation protocol. The qubit is first initialized to the $\ket{1}$ state. Then the qubit interacts with the environment for 10~$\mu$s to populate the TLS. The qubit is reset to the $\ket{0}$ state. Subsequently, the qubit and the TLS interact from 10 to 500~$\mu$s to exchange energy. Finally, the qubit $\ket{1}$ state population $P_1$ is recorded. For each value of $\Gamma_{\rm 1,TLS}$, we simulate the time evolution of $P_1$ and extract the maximum population $P_{\rm 1,max}$. The parameters used for the simulations are listed in Table~\ref{tab:simulation_para}. We repeat the simulations for different values of the unknown TLS parameters, namely, the dephasing time $T_{2,{\rm TLS}}$ and dipole moment $\boldsymbol{d}$. The corresponding $\Gamma_{\rm 1,TLS}$ threshold for the visibility depends on these parameters.

\begin{table}[t]
\caption{Simulation parameters.}
\label{tab:simulation_para}
\centering
\begin{tabular}{cc}
\hline\hline
name & values \\
\hline
$\mathbf{E}$  & 1e3~V/m  \\
$\mathbf{d}$  & 0.5, 1 D \cite{tlsweeden2025statistics}  \\
$T_{\rm bath}$  & 20$~mK$  \\
$T_{\rm 1,q}$  &  $1, 10, 100, 1000~\mu$s  \\
$T_{\phi,\rm TLS}$  & $10, 50$~ns  \\
$\gamma$   & 1~eV  \\
$\rho$  & 2000~kg/m$^3$  \\
$v_{\ell}$  & 5800~m/s  \\
$v_{t}$  & 3800~m/s \cite{yu2004study} \\
$\Delta_0$ & 10$^{-3}$ $\times f$ to 1 $\times f$ \\
\hline\hline
\end{tabular}
\end{table}

These results show that, for a given material, there is an optimal $\Gamma_{\rm 1, TLS}$ window to detect the long-lived TLSs. When $\Gamma_{\rm 1, TLS}$ is too large, the TLS energy relaxes to the environment before flowing back to the qubit. When $\Gamma_1$ is too small, the visibility is reduced because the corresponding coupling between the qubit and the TLSs is too weak. Moreover, the simulation predicts that, as qubit $T_1$ increases, the qubit re-excitation due to the~long-lived~TLSs increases.

\bibliography{references}